\documentclass[a4paper,twoside]{article}
\usepackage{tikz}
\usepackage{hyperref}   
\usepackage{enumitem}   
\usepackage{placeins}   
\usepackage{epsfig}
\usepackage{subcaption}
\usepackage{calc}
\usepackage{amssymb}
\usepackage{amstext}
\usepackage{amsmath}
\usepackage{amsthm}
\usepackage{multicol}
\usepackage{pslatex}
\usepackage{apalike}
\usepackage{algorithm2e}
\usepackage[bottom]{footmisc}
\usepackage{SCITEPRESS}     

\definecolor{figure_green}{HTML}{66C2A5}
\definecolor{figure_orange}{HTML}{FC8D62}
\definecolor{figure_blue}{HTML}{8DA0CB}
\definecolor{figure_purple}{HTML}{E78AC3}
\definecolor{figure_light_grey}{HTML}{C4C4C4}
\definecolor{figure_dark_grey}{HTML}{48494A}

\DeclareRobustCommand{\square}[2]{
\tikz[baseline=(char.base)]
    \node[shape=rectangle, fill=#2, draw=#2, text=white, inner sep=1pt] (char) {#1};\!
    }
\DeclareRobustCommand{\circlepurple}[1]{
\tikz[baseline=(char.base)]
    \node[shape=circle, fill=figure_purple, draw=figure_purple, text=white, inner sep=1pt] (char) {#1};\!
    }
\DeclareRobustCommand{\circleorange}[1]{
\tikz[baseline=(char.base)]
    \node[shape=circle, fill=figure_orange, draw=figure_orange, text=white, inner sep=1pt] (char) {#1};\!
    }
\DeclareRobustCommand{\circleblue}[1]{
\tikz[baseline=(char.base)]
    \node[shape=circle, fill=figure_blue, draw=figure_blue, text=white, inner sep=1pt] (char) {#1};\!
    }
\DeclareRobustCommand{\circledgrey}[1]{
\tikz[baseline=(char.base)]
    \node[shape=circle, fill=figure_light_grey, draw=figure_light_grey, text=figure_light_grey, inner sep=1pt] (char) {#1};\!
    }
\DeclareRobustCommand{\squaregrey}[1]{
\tikz[baseline=(char.base)]
    \node[shape=rectangle, fill=figure_dark_grey, draw=figure_dark_grey, text=figure_dark_grey, inner sep=1pt] (char) {#1};\!
    }

\begin{document}

\title{\emph{HoloGraphs}: An Interactive Physicalization for Dynamic Graphs}

\author{\authorname{
Daniel Pahr\sup{1}\orcidAuthor{0000-0001-7313-3056},
Henry Ehlers\sup{1}\orcidAuthor{0000-0002-5994-1492}, and
Velitchko Filipov\sup{1}\orcidAuthor{/0000-0001-9592-2179}
}
\affiliation{\sup{1}TU Wien, Austria}
}


\abstract{%
We present \emph{HoloGraphs}, a novel approach for physically representing, explaining, exploring, and interacting with dynamic networks. \emph{HoloGraphs} addresses the challenges of visualizing and understanding evolving network structures by providing an engaging method of interacting and exploring dynamic network structures using physicalization techniques.
In contrast to traditional digital interfaces, our approach leverages tangible artifacts made from transparent materials to provide an intuitive way for people with low visualization literacy to explore network data. The process involves printing network embeddings on transparent media and assembling them to create a $3D$ representation of dynamic networks, maintaining spatial perception and allowing the examination of each timeslice individually. Interactivity is envisioned using optional \textit{Focus+Context} layers and overlays for node trajectories and labels. Focus layers highlight nodes of interest, context layers provide an overview of the network structure, and global overlays show node trajectories over time. In this paper, we outline the design principles and implementation of \emph{HoloGraphs} and present how elementary digital interactions can be mapped to physical interactions to manipulate the elements of a network and temporal dimension in an engaging matter. We demonstrate the capabilities of our concept in a case study. Using a dynamic network of character interactions from a popular book series, we showcase how it represents and supports understanding complex concepts such as dynamic networks. 
}

\onecolumn \maketitle \normalsize \setcounter{footnote}{0} \vfill

\begin{figure*}
    \centering
    \includegraphics[width = \textwidth]{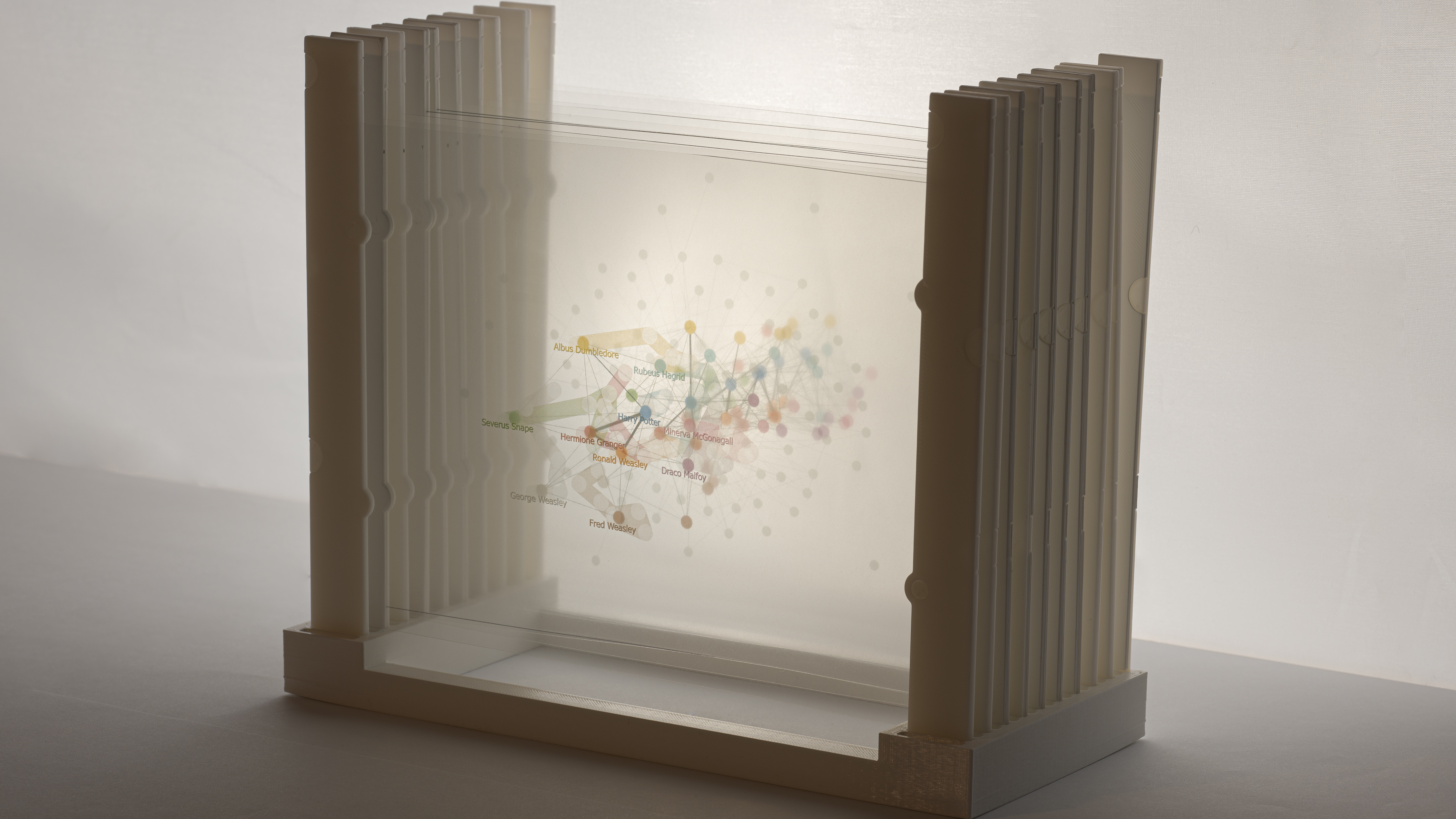}
    \caption{A \textit{HoloGraph}. We display a dynamic graph by producing and embedding the individual timeslices, printing them on transparent media, and arranging them equally spaced. An overlay shows interesting nodes' trajectories over time and per individual timeslice.}
    \label{fig:teaser}
\end{figure*}

\section{\uppercase{Introduction}}

Dynamic network visualization aims to support extracting insights and making sense of dynamically changing network structures \cite{beck_taxonomy_2017}.
Subsequently, dynamic graphs and their visualization have found common use across a variety of domains, from the social sciences \cite{oelke_fingerprint_2013}, through software engineering \cite{rufiange_animatrix_2014}, to metabolic pathway analysis \cite{rohrschneider_visual_2010}.
Several approaches to visualizing such graphs have been proposed, such as animated node-link diagrams \cite{hayashi_initial_2013}, layered matrix representations \cite{vehlow_radial_2013}, integrated $1.5D$ representations \cite{shi_15d_2015}, or a superimposition of multiple $2D$ embeddings from different points in time~\cite{filipov_timelighting_2023}. With multiple embeddings for each point in time, we can produce \textit{space-time cubes} ($2D + t$ or $3D$), that encode the temporal aspect in a third dimension.
However, virtual $3D$ representations suffer from occlusions, distortions, and parallax effects \cite{bach_graphdiaries_2014}. 
This can, in turn, obscure the graph's topology and change over time, negatively impacting user perception and understanding.
In contrast, by arranging the slices side-by-side, i.e. juxtaposing them, we lose the ability to perceive nodes' movement over time between the individual timeslices \cite{beck_taxonomy_2017,ehlers_visualizing_2024,filipov_network_2023}.

A key benefit of data physicalization is that we can examine $3D$ representations tangibly and intuitively by physically manipulating them. 
Research shows that certain drawbacks of $3D$ representations on screens can be overcome by such physicalizations~\cite{jansen_evaluating_2013}.
Specifically, research regarding networks in physical space has focused on both $2D$ embeddings with additional tactile encodings~\cite{drogemuller_haptic_2021,drogemuller_towards_nodate} as well as $3D$ embeddings with added interactivity~\cite{bae_computational_2024}.
Drogemuller et al.\cite{drogemuller_haptic_2021} present their findings about physicalized graphs being more engaging and fun to interact with, compared to their screen-based counterpart.

Given these outcomes regarding engagement and fun, it is unsurprising that interactive data physicalizations have value particularly related to education and visualization literacy \cite{omalley_literature_2004}.
Manipulable physical representations have great value in educational settings, conceptually allowing for greater engagement, understanding, and learning \cite{jansen_opportunities_2015,jansen_interaction_2013,pahr_investigating_2024}. 
We argue that there still are numerous opportunities to leverage the unique strengths of data physicalization to communicate complex scientific phenomena and concepts, such as dynamic graphs and space-time cubes.


We introduce a novel workflow to create interactive, physical representations of dynamic networks.
We make this workflow accessible by moving the focus away from complex $3D$-printing technologies and toward widely and cheaply available materials that can be assembled with various means. 
By printing network embeddings on overhead projector slides, we obtain transparent slices that can be assembled in parallel to create a $3D$ appearance: a \textit{HoloGraph}. 
This allows us to keep the spatial perception of the \textit{space-time cube} while enabling the examination of individual timeslices at any point.
We divide each of the network's embeddings into \textit{focus slices} that highlight and track nodes of interest and \textit{context slices} that depict the contextual structure of the network at different points in time. 
We also provide separate global slices such as \textit{label overlays} for nodes of interest and \textit{trajectory overlays} to track nodes' movement over time.
\href{https://osf.io/4u2e9/?view_only=751235378e564086beee9de8d37a6686}{The codebase, printable versions of our networks, as well as $3D$-printable meshes we used for our sculptures, can be found online}\footnote{\url{https://osf.io/4u2e9/?view_only=751235378e564086beee9de8d37a6686}}.
We present a demonstration of our approach's utility in a case study, exploring the evolving relationships of characters across books of the ``\textit{Harry Potter}'' series.
In summary, the \textbf{contributions} of this work are two-fold:
\begin{itemize}[noitemsep]
    \item We present the development of a novel workflow with which to create physical and interactive representations of dynamic graphs.
    \item We demonstrate the utility of \textit{HoloGraphs} in a case study to highlight the value and importance of engagement when learning about concepts such as dynamic graphs.
\end{itemize}

\section{Related Work}

\paragraph{Data Physicalization} Data physicalization transforms abstract data into tangible forms, leveraging the physical properties of artifacts and materials to encode information~\cite{jansen_opportunities_2015}. This approach has shown promise in making data more accessible and engaging. For instance, $3D$ physical representations can outperform their screen-based counterparts in efficiency for certain tasks~\cite{jansen_evaluating_2013}. Furthermore, they can also enhance memory retention and engagement compared to virtual approaches~\cite{stusak_evaluating_2015,hurtienne_move_2020}.
Data physicalization approaches often include interactive elements (``hands-on''), 
allowing users to engage with the data actively, performing common tasks such as filtering and selecting interesting data items~\cite{brehmer_multi-level_2013}. For instance, \textit{Vol2Velle}~\cite{stoppel_vol2velle_2017} allows the selection of transfer function parameters by rotating disks. Similarly, in \textit{Volograms}~\cite{pahr_vologram_2021} individual slices can be removed to be inspected individually or to examine otherwise obstructed regions. Schindler et al.~\cite{schindler_anatomical_2020} propose the use of color filters to allow the filtering of different anatomical regions. Bae et al.~\cite{bae_computational_2024} propose a pipeline for creating network physicalizations that allow simple interactions with the nodes of the network using electrical circuitry.

\paragraph{Physicalizing Networks} 
Networks are powerful structures used to model and visualize data as a set of entities and relationships between them. They are widely employed in different domains to help understand connections, detect patterns, and identify influential actors~\cite{lee_task_2006,jae-wook_ahn_task_2014}. Physicalizing networks adds a tactile dimension to the data exploration and analysis process, enhancing accessibility for the visually impaired individuals~\cite{drogemuller_towards_nodate,mcgookin_clutching_2010}. Recent research investigates how combining visual and haptic exploration of physical node-link diagrams benefits the understanding of such structures~\cite{drogemuller_haptic_2021} and how physicalized networks can support a better spatial perception through tangible interactions~\cite{mcguffin_path_2023}.
There is growing interest in exploring $3D$ immersive environments for network visualization~\cite{kotlarek_study_2020,sorger_immersive_2019,oh-hyun_kwon_study_2016,colin_ware_visualizing_2008}. Such novel representations and interaction techniques promise new ways to engage with the data. Integrating dynamic network visualization with data physicalization offers a tangible, interactive method to explore complex concepts such as evolving network structures, and behavior over time, and identifying patterns and trends.
Existing physicalizations of networks often focus on simple, static structures. However, dynamic networks introduce an additional layer of complexity, considering the temporal evolution of the entities and their relationships; an aspect that is crucial for understanding behavior over time~\cite{jae-wook_ahn_task_2014}. Approaches typically tackle this problem by aggregating (or timeslicing) the network's temporal dimension~\cite{archambault_animation_2011,bach_descriptive_2017}, while continuous representations (event-based) capture changes occurring at finer temporal granularities~\cite{simonetto_event-based_2018,arleo_event-based_2022}. 

\paragraph{Accessible Fabrication} Digital fabrication, i.e. $3D$ printing technology, is one of the most common approaches to creating $3D$ artifacts~\cite{djavaherpour_data_2021}. However, more affordable methods exist that can make them accessible to a broader audience. Stoppel and Bruckner~\cite{stoppel_vol2velle_2017}, for example, use transparent disks to create interactive volume visualizations, while Pahr et al.~\cite{pahr_vologram_2021} present hologram-like structures from segmented volumetric data. Raidou et al.~\cite{raidou_slice_2020} demonstrate how volumetric data can be printed on transparent material to create $3D$ sculptures without the use of sophisticated technology and Schindler et al.~\cite{schindler_nested_2022} show methods for creating nested paper structures for anatomical education. Such approaches do not require expensive hardware or technology, resulting in more affordable and accessible physical data representations.

\section{HoloGraphs}
\label{sec:holographs}

\begin{figure*}[ht!]
    \begin{subfigure}[t]{0.24\linewidth}
        \centering
        \includegraphics[width=0.9\linewidth]{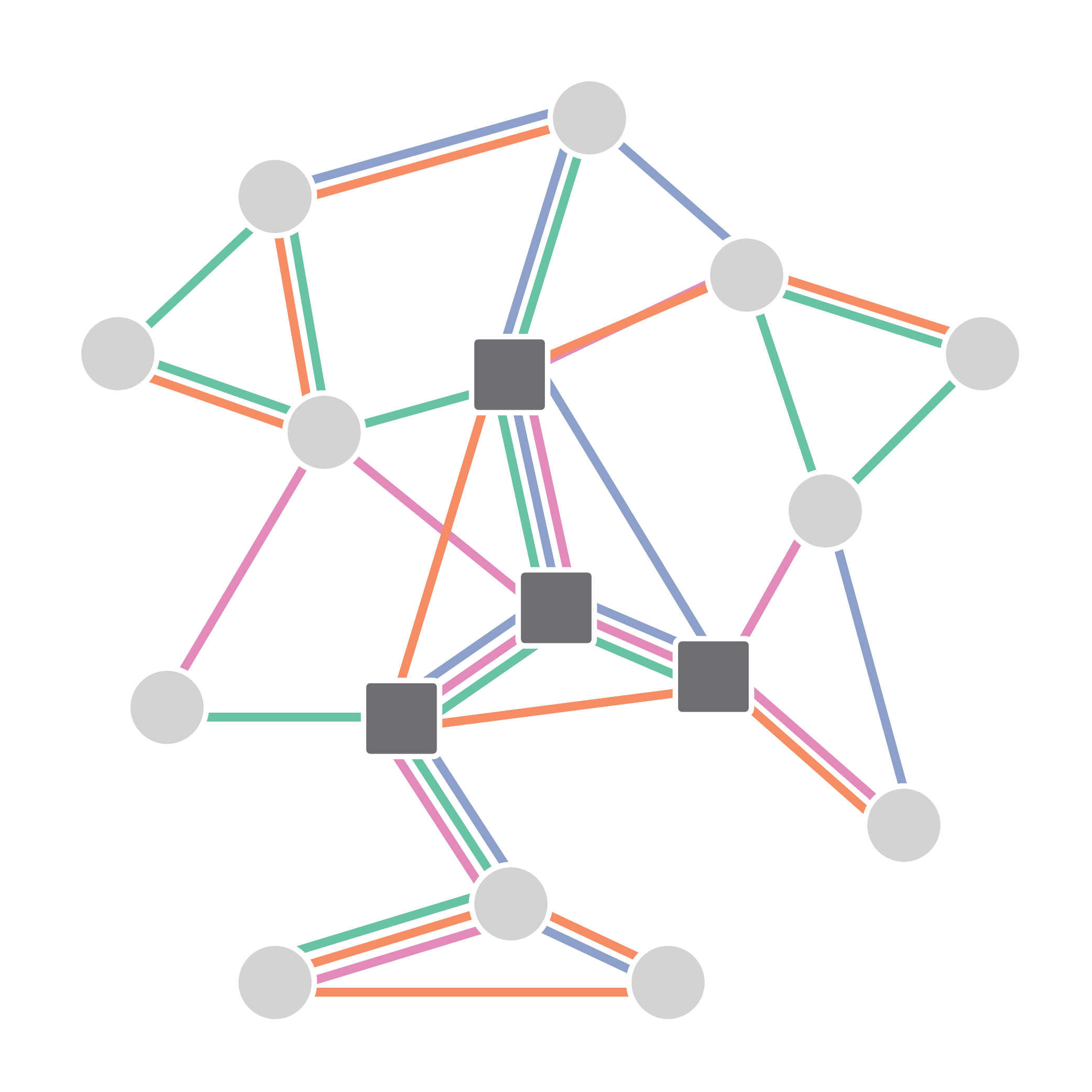}
        \caption{Super Graph}
        \label{fig:pipeline:super_graph}
    \end{subfigure}
    \begin{subfigure}[t]{0.24\linewidth}
        \centering
        \includegraphics[width=0.9\linewidth]{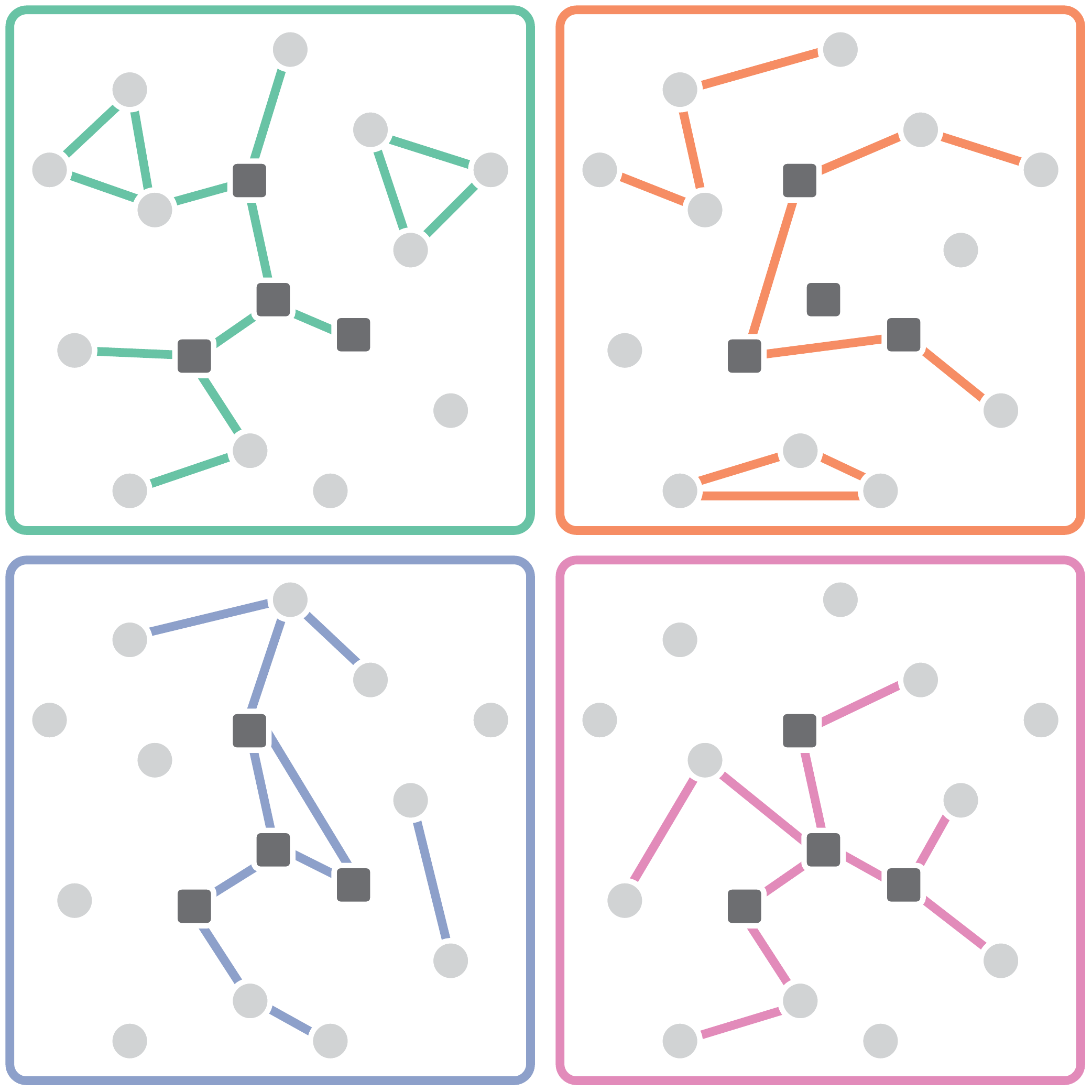}
        \caption{timeslice Subgraphs}
        \label{fig:pipeline:timeslice_graphs}
    \end{subfigure}
    \begin{subfigure}[t]{0.24\linewidth}
        \centering
        \includegraphics[width=0.9\linewidth]{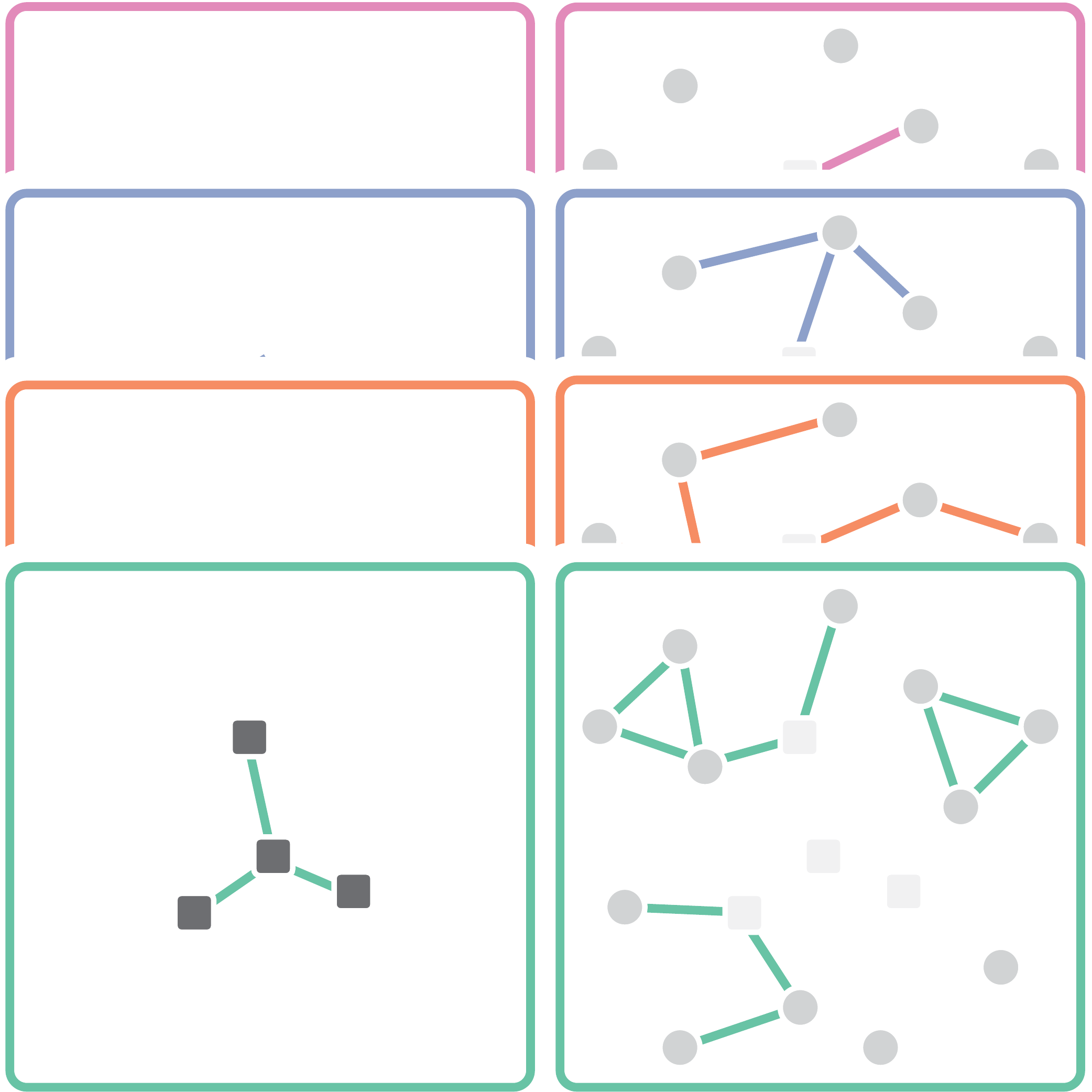}
        \caption{Focus+Context Subgraphs}
        \label{fig:pipeline:focus_and_context_graphs}
    \end{subfigure}
    \begin{subfigure}[t]{0.24\linewidth}
        \centering
        \includegraphics[width=0.9\linewidth]{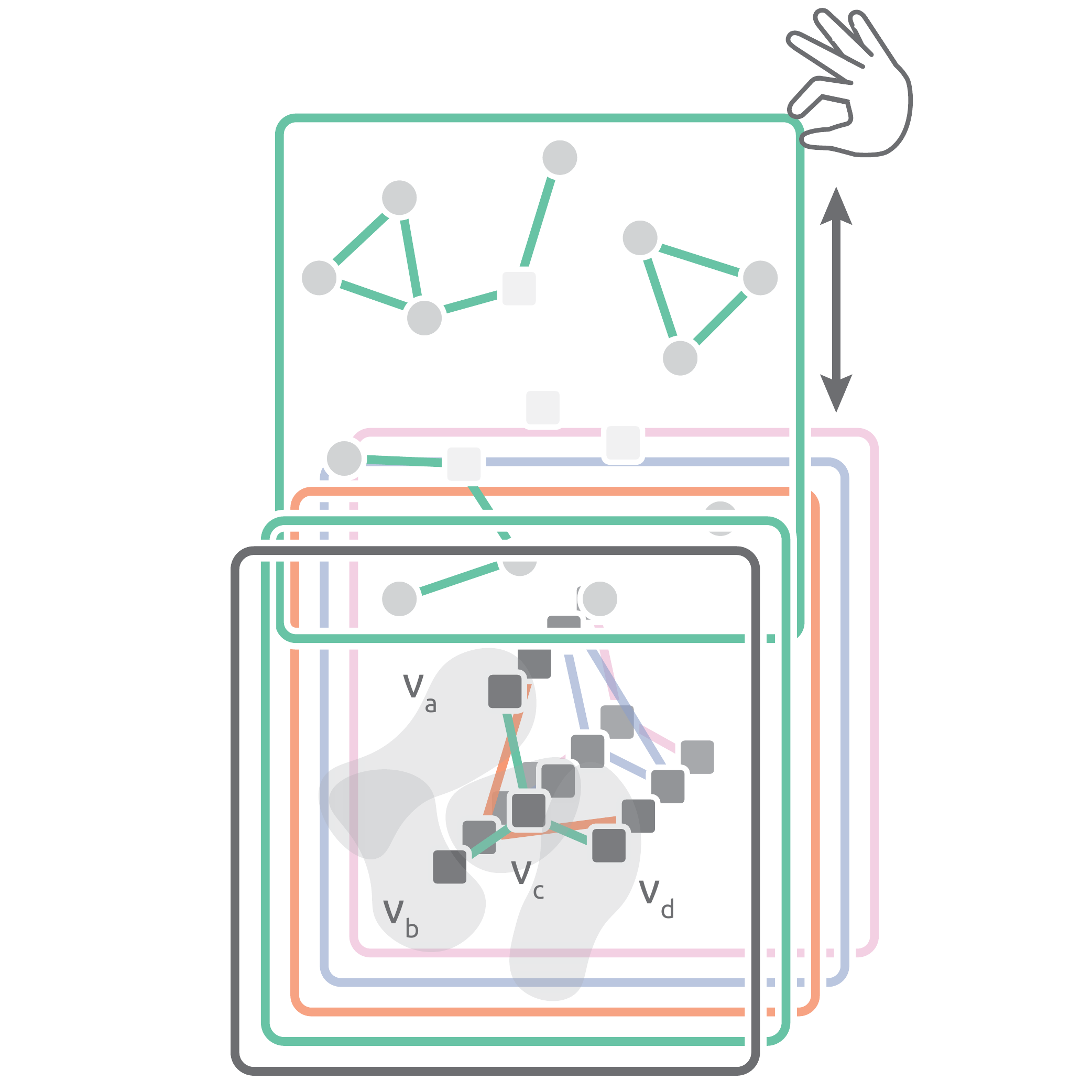}
        \caption{HoloGraph}
        \label{fig:pipeline:holo_graph}
    \end{subfigure}
    \caption{    
A dynamic graph (a), where connections between nodes and links differ between different points in time, is split up into timeslices \square{$t_{1}$}{figure_green}, \square{$t_{2}$}{figure_orange}, \square{$t_{3}$}{figure_blue}, \square{$t_{4}$}{figure_purple}, representing the state of the network at different points in time (b). 
To emphasize nodes of interest (\squaregrey{x}), we divide the timeslices into focus (left) and context (right) subgraphs (c).
Arranging the slices in parallel creates a \textit{space-time cube} appearance (d). Individual timeslices can be removed for inspection and global overlays show the focus nodes' movements over time.
For illustration purposes and simplicity, each timeslice subgraph shares the same layout. In practice, each timeslice subgraph is laid out semi-independently of the others, resulting in node movement between time points.
    }    
        \label{fig:pipeline}
        \vspace{-10pt}
\end{figure*}

The physicalization of node-link diagrams is conventionally done using $3D$-printing techniques~\cite{bae_computational_2024}. 
However, such representations limit the number of nodes that can be printed and hence displayed owing to the time and cost associated with 3D printing; a limitation we here aim to side-step.
Here, Pahr et al.~\cite{pahr_vologram_2021} propose methods to process and (interactively) display medical volumetric data using regular printers and transparent printable media. 
Moreover, outside of the context of data physicalization, Filipov et al.~\cite{filipov_timelighting_2023} describe a method to visualize dynamic networks by projecting the \textit{space-time cube} embedding to a $2D$ representation. 
This $2D$ representation is an orthogonal projection of the network's topology over time in order to visualize node trajectories and behavior. 
Here, combining these two approaches, we propose to show the individual timeslices created by such a method in a $3D$ environment by printing each time-slice-subgraph embedding on transparent media and arranging them as parallel slices. To add interactivity, we propose the use of various removeable \textbf{overlays} with which to display added \textbf{context}, draw \textbf{focus}, and provide \textbf{global characteristics} of the data.
Figure~\ref{fig:pipeline} provides a simplified overview of the proposed process.

\paragraph{Definitions.}

Given a dynamic graph, formulated here as a time-sliced graph 
$G = (V_{S},E_{S},T)$, where
$V$ the total set of nodes and 
$E$ the total set of undirected links
across a set of timeslices $T$. 
For some timeslice $t \in T$ node-time-slice pairs 
$(v_{a}, t)$ and
$(v_{b}, t)$ in $V_S$ 
are connected by links
$E_{S} \subseteq V_{S} \times V_{S}$, where 
$V_{S} \subseteq V \times T$. 
For a set of \textbf{focus nodes} $V_{F} \subseteq V$, we aim to highlight their position in the different timeslices over a disjoint set of \textbf{context nodes} $V_{C} \subseteq V$.
First, the dynamic \textbf{super graph} $G$ (Figure~\ref{fig:pipeline:super_graph}) is separated into its constituent \textbf{time-sliced subgraphs} $G_s$ (Figure~\ref{fig:pipeline:timeslice_graphs}).
These individual subgraphs are then each further broken up into their individual \textbf{\textit{Focus+Context} subgraphs}, $G_{sF}$ and $G_{sC}$
respectively, such that 
    $E_{sF} = V_{F}  \times V_{F}$, as well as
    $E_{sC} = V_{s} \times  V_{C}$ (Figure~\ref{fig:pipeline:focus_and_context_graphs}).
These \textit{Focus+Context} subgraphs are then individually printed and mounted in a physical rack, forming (together with an additional labeled overlay that is printed along nodes' time-dependent trajectories) a tactile and interactive dynamic graph physicalization: the \textbf{HoloGraph} (Figure~\ref{fig:pipeline:holo_graph}).

\paragraph{Virtual Embedding}
\begin{figure}[b!]
    \centering
         \includegraphics[width = \linewidth]{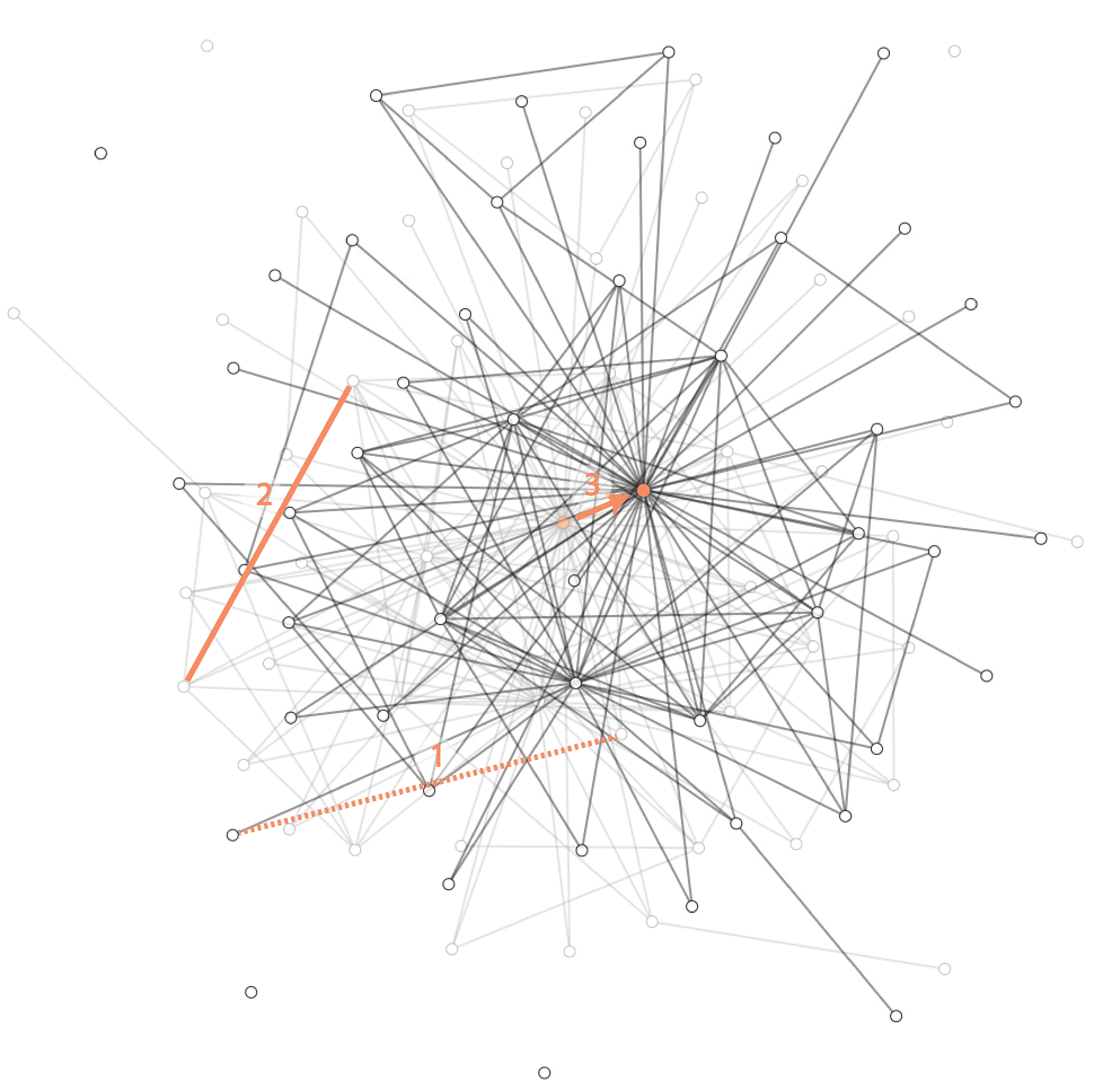} 
    \caption{Embeddings for individual timeslices overlaid. The embedding of the subsequent slice is dependent on the previous. Disappearing (1) and appearing (2) links and the subsequent re-embedding causes movement of the nodes (3).}
    \label{fig:embeddings}
\end{figure}
The \textbf{layout} of the network is computed using \texttt{D3.js}'s particle-based force-directed layout algorithm~\cite{bostock_d_2011}.
Here, the key challenge of laying out a dynamic network is to strike a \textbf{balance between having enough change} in node placement between timeslices to effectively reflect the network's evolution over time, while also restricting said movement sufficiently in order to ensure \textbf{layout and network structure are still preserved}. 
Brandes et al.~\cite{brandes_visualization_2012} present several strategies for doing so, most notably \textit{aggregation} and \textit{anchoring}. In \emph{HoloGraphs} we make use of an \textbf{anchoring} approach to computing the layout of the dynamic network: 
for each timeslice $t$, we utilize the layout of the previous timeslice $t-1$ as an initial layout before commencing the layout process. 
This ensures that the nodes' \textbf{movement over time will remain consistent} (i.e no flickering or popup effects) while preserving the viewer's mental map in the transitions from timeslice to timeslice~\cite{hutchison_mental_2013}. 

\begin{figure}[hb!]
    \centering
        \begin{subfigure}[t]{0.95\linewidth}
        \centering
        \includegraphics[width=\linewidth]{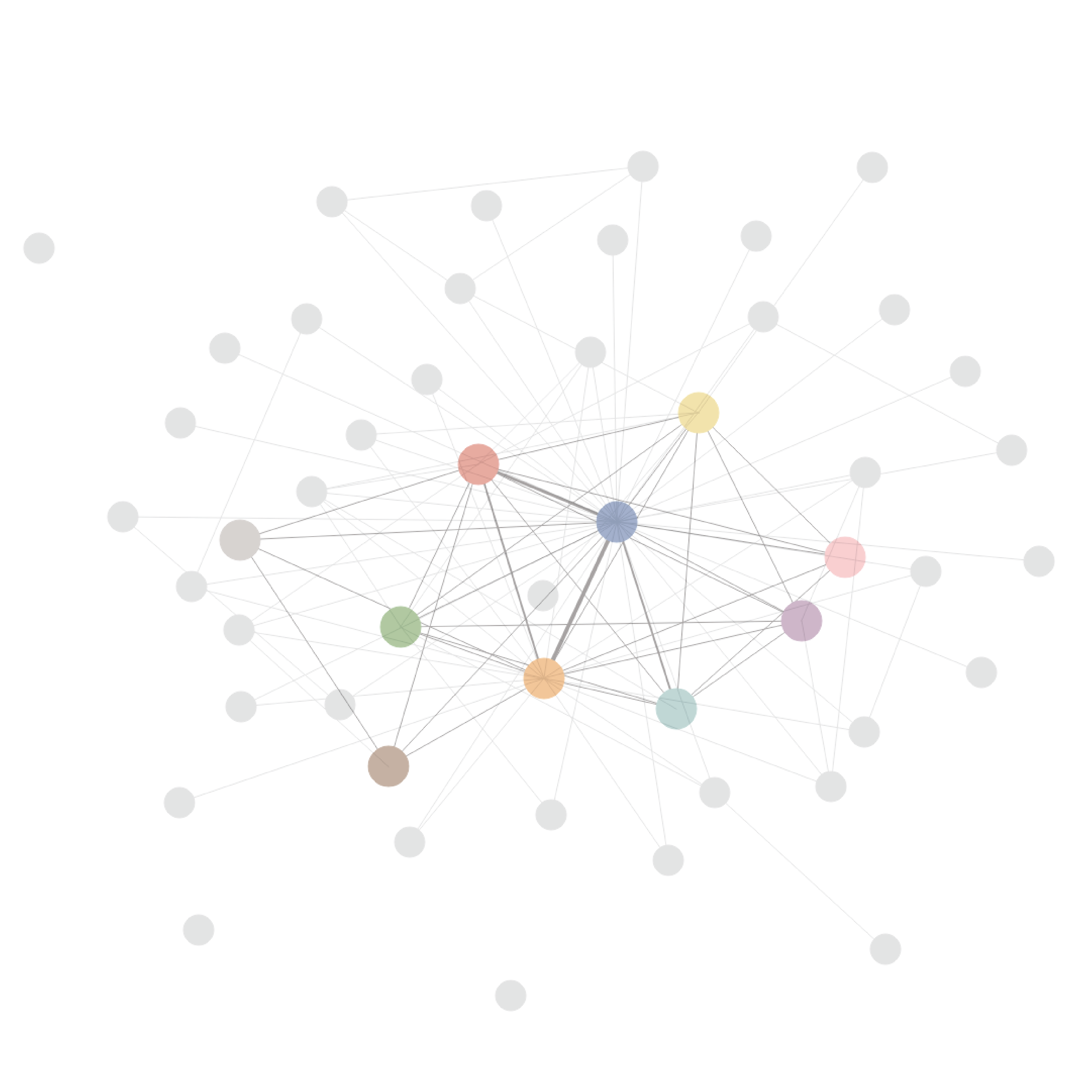}
        \caption{Focus + Context}
        \label{fig:phys_embed:focus_context}
    \end{subfigure}
        \centering
        \begin{subfigure}[t]{0.95\linewidth}
        \centering
        \includegraphics[width=\linewidth]{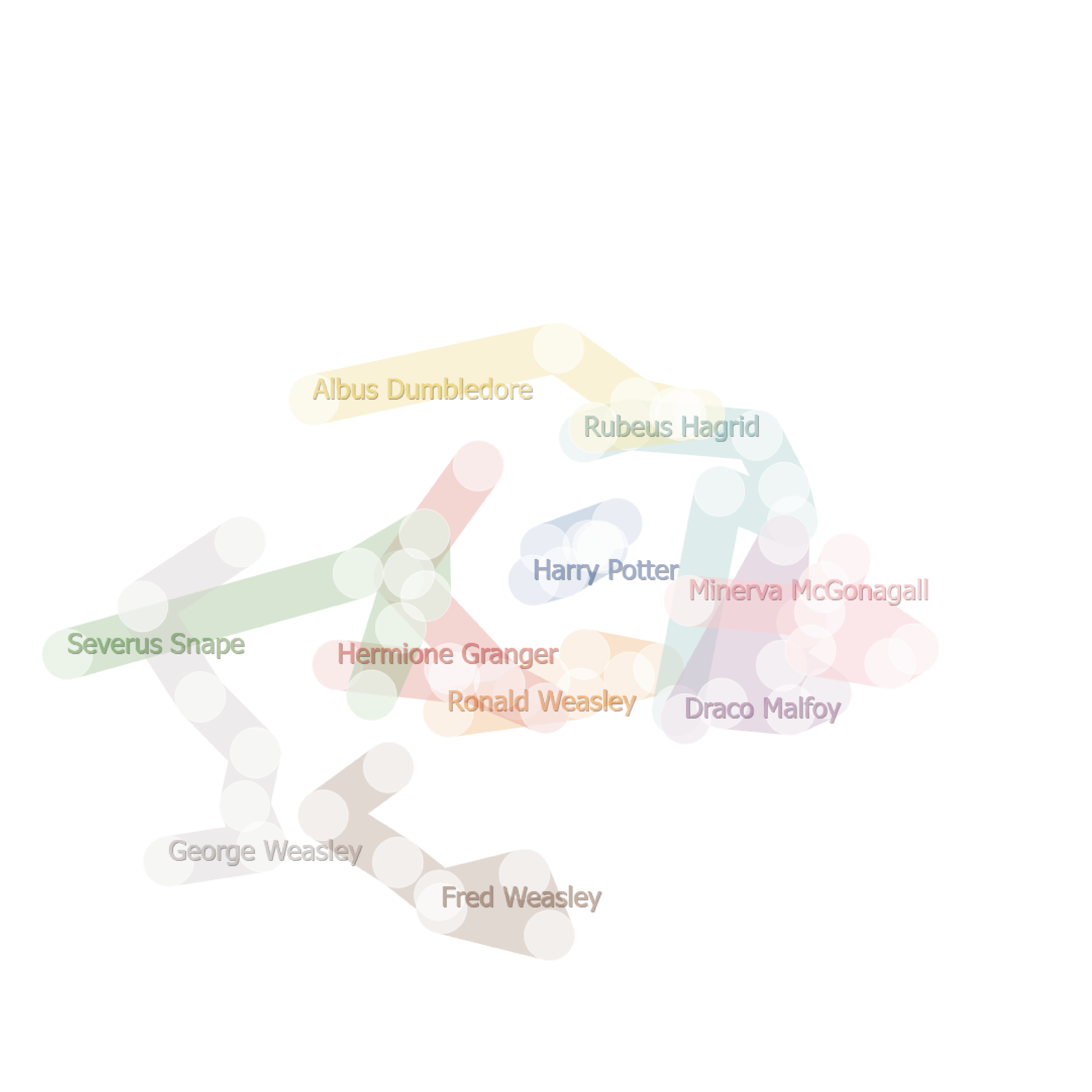}
        \caption{Global overlays}
        \label{fig:phys_embed:global}
    \end{subfigure}
    \caption{Physical embeddings for Focus + Context slices as well as Labels + Trajectories overlays. (a) shows a superimposition of the focus subgraph, indicated by larger and colored nodes, and the context subgraph with smaller, faint grey nodes for a single timeslice. (b) shows the global overlays for focus node trajectories and labels.}
    \label{fig:phys_embeddings}
\end{figure}
Here, the movement of the nodes between timeslices depicts their trajectories over time as edges form or dissolve, pulling or pushing the nodes, respectively, reflecting the network's evolving structure. 
In contrast, an aggregation approach, i.e. computing singular positions for each node based on the layout of the super graph (Figure \ref{fig:pipeline:super_graph}), would result in no node movement over time. 
Figure~\ref{fig:embeddings} shows an example of the layouts of multiple timeslices overlaid on top of each other and highlighting nodes' movement over time (orange lines), making it difficult to make sense of changes over time. 
Instead, the focus nodes' trajectories, i.e. their movement between timeslices, can be better perceived in Figure~\ref{fig:phys_embed:global}.

\paragraph{Physical Embedding}

\begin{figure*}[t!]
    \centering
        \begin{subfigure}[t]{0.32\linewidth}
        \centering
        \includegraphics[width=\linewidth]{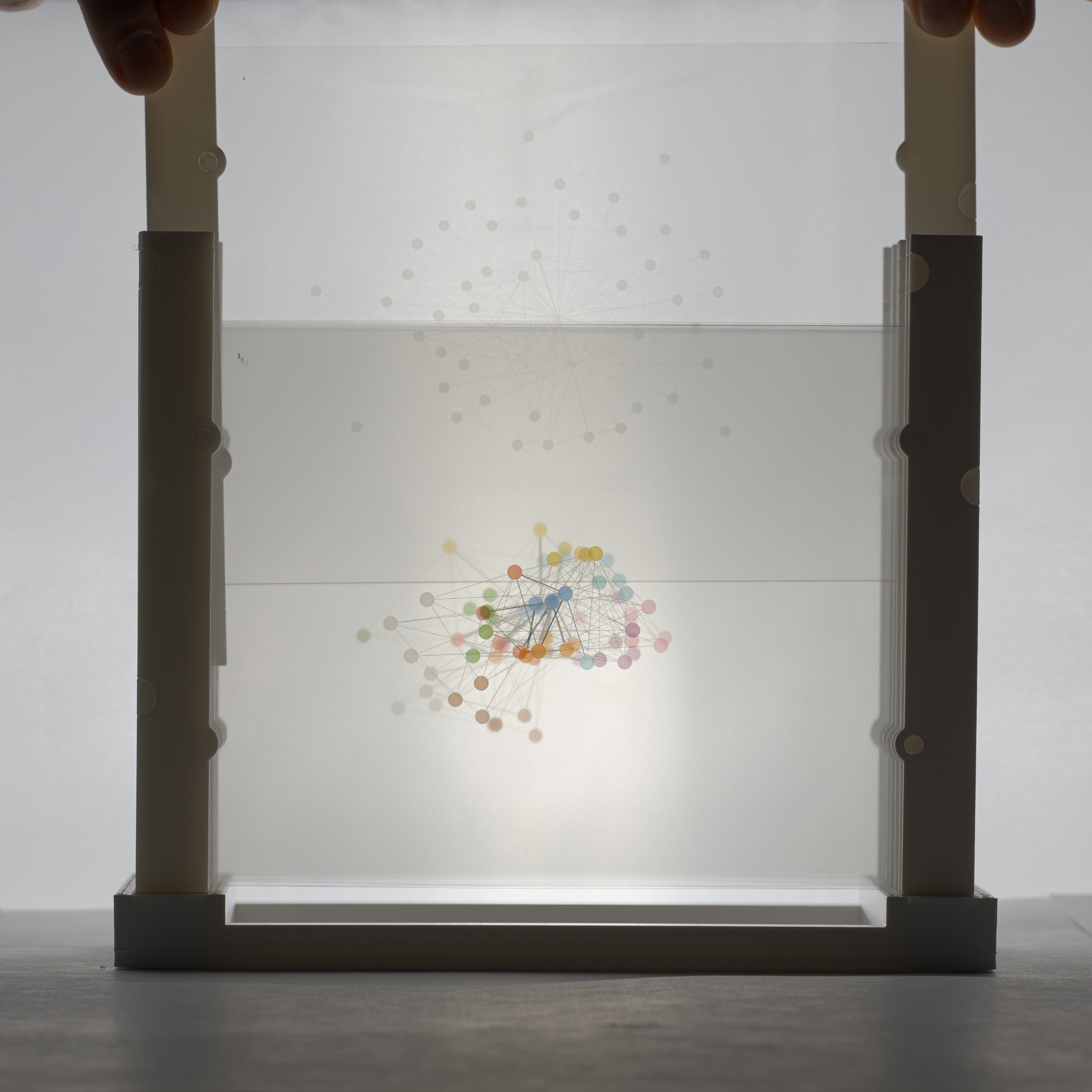}
        \caption{Context slices}
        \label{fig:interactive:focus_context}
    \end{subfigure}
    \hfill
    \begin{subfigure}[t]{0.32\linewidth}
        \centering
        \includegraphics[width=\linewidth]{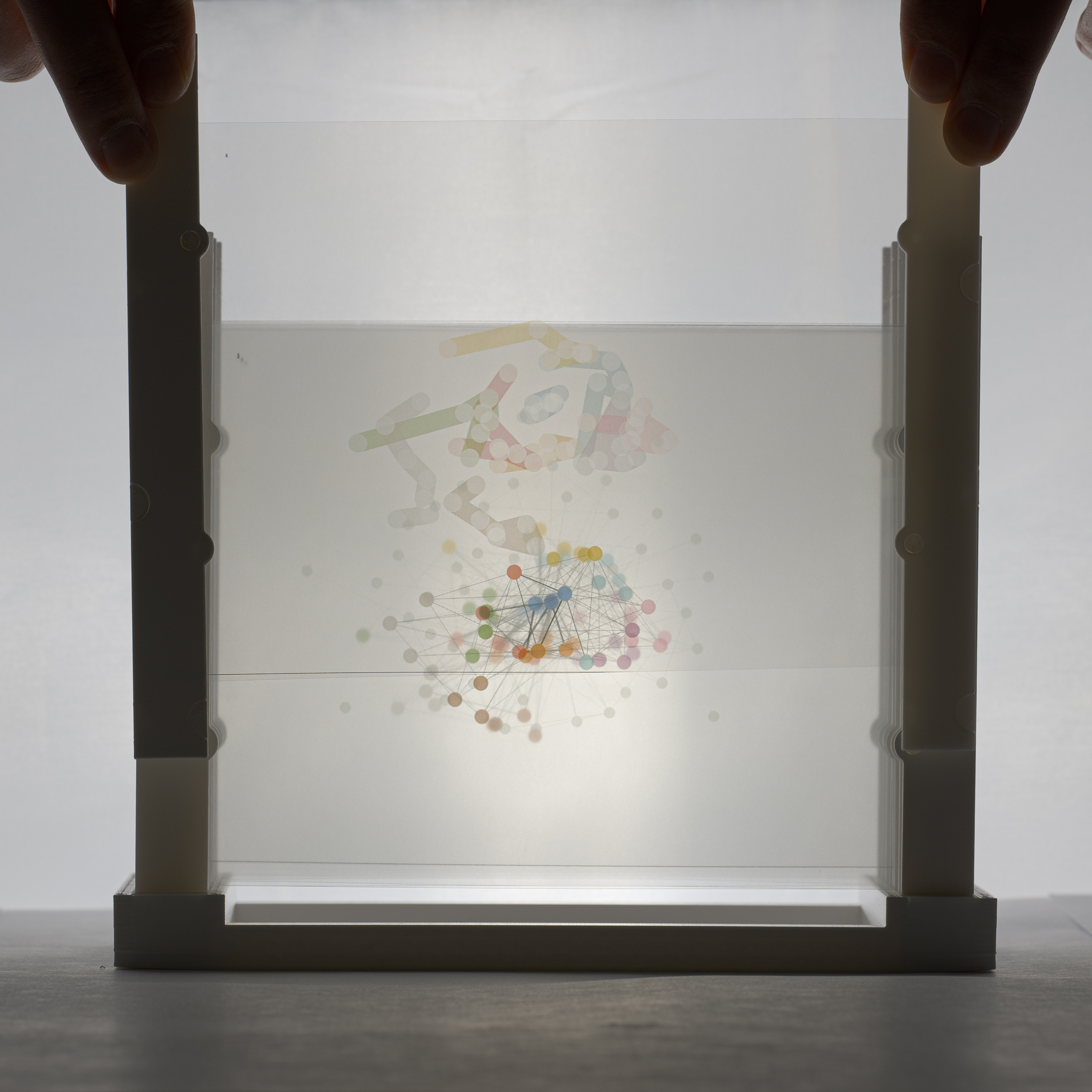}
        \caption{Trajectory overlay}
        \label{fig:interactive:trajectories}
    \end{subfigure}    
    \hfill
    \begin{subfigure}[t]{0.32\linewidth}
        \centering
        \includegraphics[width=\linewidth]{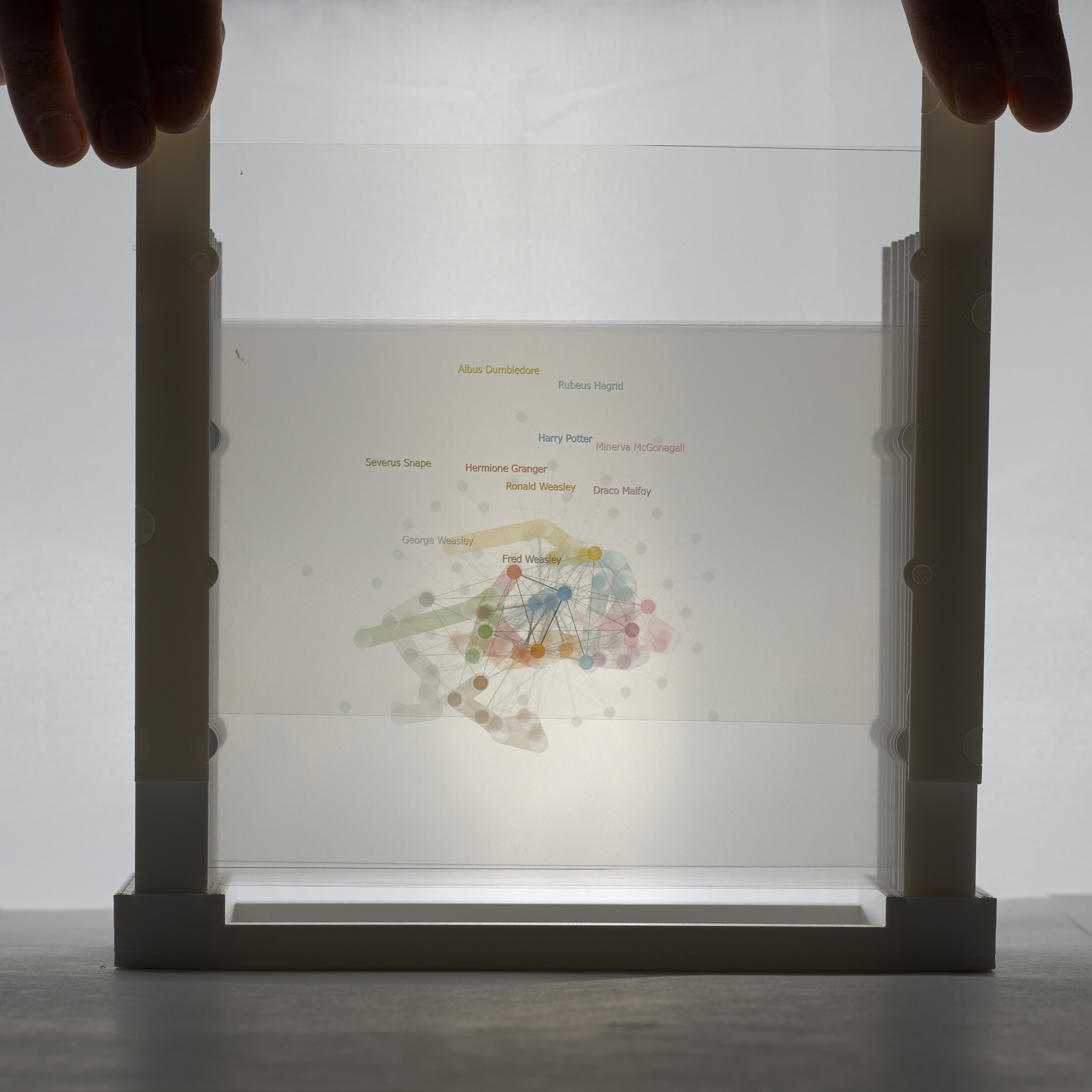}
        \caption{Labels}
        \label{fig:interactive:labels}
    \end{subfigure}    
    \caption{
        Composition of a \textit{HoloGraph}. 
        Focus slices show the nodes of interest at every timeslice. 
        Context slices can be added to each focus slice individually (a). 
        Node trajectories of focus nodes can be added as a global overlay (b), together with focus node labels (c). 
    }
    \label{fig:interactive}
\end{figure*}
Figure~1 shows our version of a slide holder, used to stack the individual timeslices printed on transparent media. 
To make sure that our method supports a wide variety of fabrication techniques, we add \textbf{manufacturing parameters} to the physical embeddings. 
The physical representation depends on the \textbf{transparency} and \textbf{format} of the slides used. Additionally, the desired \textbf{colors} can be chosen depending on the availability of color printing.

First, we transform the node positions of the individual timeslices for the physical embedding to the format of the desired output. This corresponds to the \textbf{paper size}, with additional \textbf{margins} that ensure that the slide holders do not obscure any data.
For all timeslices, we calculate the minimum and maximum $x$ and $y$ positions of the nodes. We then map the values to the space between the chosen margins.
We then create \textbf{separate embeddings} for the chosen \textbf{focus} and \textbf{context} subgraphs. 
To emphasize nodes of interest, the focused subgraph is embedded using distinct node colors. 
In the context subgraph, we use smaller node diameters and a lighter hue of gray to avoid visual obstruction in the \textit{HoloGraph}.
The \textbf{edge weights} are encoded into edge thickness in both subgraphs. 
Figure~\ref{fig:phys_embed:focus_context} shows both of these embeddings superimposed. Printing the focus and context embeddings separately allows users to customize the representation interactively. Figure~\ref{fig:interactive:focus_context} shows the insertion or removal of a context slice from a \textit{HoloGraph}.
Finally, we also create \textbf{global overlays} to \textbf{trace the movement} of focus nodes over time, shown in Figure~\ref{fig:phys_embed:global}. 
For this, we draw a polyline for each focus node through all its positions in the timeslices. At the node positions, we add a bright circle to emphasize where the node can be found. Figure~\ref{fig:interactive:trajectories} shows the process of adding this slice to the \textit{HoloGraph}. We create a separate overlay for \textbf{node labels}. This way the labels can be removed freely, minimizing possible obstructions as shown in Figure~\ref{fig:interactive:labels}. 


\section{Case Study}
\label{sec:case}

To examine the potential of \textit{HoloGraphs}, we present a case study highlighting social network interactions between characters in a famous children's book series.

\paragraph{Data}
We use a dynamic network representing the character interactions in the series of Harry Potter books by J.K. Rowling\footnote{\url{https://github.com/nikhil-ravi/harry-potter-interactions}} to show the capabilities of our approach. The dataset is processed as a time-sliced network, where the nodes represent the characters in the books, and the weighted edges represent the number of interactions between those characters within each book. 
The individual timeslices represent the state of the character interaction graph within each book. We reduced the size of the original dataset to visualize the most central and influential actors as well as the most important relationships. 
By filtering out the links whose weights were not in the top 10 percentile we additionally reduce clutter and put an emphasis on the more important relationships between the characters of the data. 
This process resulted in a set of 111 nodes (characters) and 612 weighted edges (relationships) over 7 timeslices (books).
From these $110$ nodes, ten focus nodes, i.e. ``egos'' \cite{ehlers2024me}, were selected (Figure \ref{fig:phys_embed:global}).
 
\paragraph{Implementation}

Our pipeline's virtual embedding was done using \texttt{d3.js}~\cite{bostock_d_2011} and its physical embedding using \texttt{qt/python}. 
For the focus nodes, we use the top 10 nodes with the highest degree centrality, present in each of the seven timeslices.
We used laser-printer-compatible overhead slides as the medium for the slices.
The slide holder and base were $3D$ printed in a Prusa MK2S.
To use the entire printer surface, we decided to print the slices on half of an A4 page, resulting in 2 slices per page.
The slices were attached to the holders using a hole punch on both sides of each slide to keep them in place.
The slice holders fit into the left and right sides of the base, ensuring slices remained parallel when mounted. 
Figure~\ref{fig:teaser} shows a fully assembled \textit{HoloGraph}, Figure~\ref{fig:interactive} shows how different overlays are added or removed. 

\paragraph{Findings}

Using our Harry Potter \textit{HoloGraph}, we present some notable findings that can be extracted from the time-sliced dynamic network.
Our participant had an interest in the Harry Potter series as well as basic knowledge of node-link diagrams. They were, however, unfamiliar with the concept of dynamic graphs.
We then asked the participant to explore our \textit{HoloGraph}, discussing notable events, character moments, and group interactions. 
We present the insights, findings, and \emph{emphasize} the reasoning and suggestions we obtained with the goal of improving our approach in communicating dynamic networks to non-expert audiences. The aim of our work is to provide affordable data physicalizations to increase knowledge and engagement using a hands-on approach and interesting datasets. 

Harry Potter, as the protagonist of all the books, takes a stable position in the center of the graph. \emph{His movements are minimal, however, we can observe how other nodes of interest behave in relation to him.}
Rubeus Hagrid, for example, is a central character in the first books, often interacting with Harry as his ``window to the wizard world'', however, in later books, he leaves the school and has other interactions. \emph{While he remains tethered to Harry and in a stable orbit around him, his other interactions cause him to move away from the central set of characters}.
Albus Dumbledore's interactions with Harry are often distant and occur at the end of the individual books. In books five and six, he becomes closer associated with Harry, and they form a personal relationship, culminating in private tuition sessions. \emph{After his departure from the active cast in the sixth book, his node moves towards new characters related to his backstory and away from the central set of characters}.
Fred and George Weasley are twins and are often shown in interactions together. This is supported by their \emph{very similar, sometimes parallel trajectories}. They \emph{become separated close to the end of the last book when Fred's trajectory is a little shorter} than Georges's.
Finally, one of the most faceted characters, Severus Snape also has an interesting trajectory. \emph{He spends most of the time in a stable position, close to the center, as part of the ``Hogwarts cluster'}'. 
He also joins another faction dedicated to the protection of Harry in the later books, leading to continued proximity. \emph{Finally, he changes factions again when his node is pulled away from the other teachers in the school and towards the antagonists}.

\paragraph{Issues}

The node trajectories often point towards shifting associations of the characters with different groups. However, in the presented version of the Harry Potter \textit{HoloGraph}, these groups are not highlighted. Our expert recommended \emph{visualizing these groups by highlighting pertaining nodes in different colors in the context view}. Node trajectories could be easier explained by having this added context.
We also omitted different characters in favor of readability. Our selection of nodes of interest was purely based on node centrality. A suggestion here was to \emph{emphasize a different selection of characters} that interact in different ways with the central characters. Notably, Voldemort, the primary antagonist of the story was not included in the focus nodes for lack of centrality. Other nodes, such as the love interests of different characters could have interesting trajectories in conjunction with the associated characters. 
Finally, while the trajectory overlay was overall helpful in finding nodes of interest in the different timeslices, it led to some confusion. Our expert thought that a certain move of the node occurred early on in the data, but noticed that the node was in a different place than expected at that time.
The participant suggested \emph{indicating a direction in the trajectory lines}, for example by narrowing the lines from first to last timeslice. 


%

\FloatBarrier
\section{Conclusions and Future Work}

\paragraph{Takeaways.} Our work presents an \textbf{accessible} and \textbf{affordable} way to create \textbf{interactive} physicalizations of dynamic graphs. Our overarching goal is for this approach to be applied to improve visualization literacy and provide an engaging method for communicating science-related concepts in education settings.
In our work, we investigated small to medium-sized graphs using the Harry Potter dataset. 
However, the number of nodes that can be displayed is limited only by printer resolution and the size of the used medium.
Because we are printing our embeddings with $2D$ printing devices, the number of nodes can be much larger than in $3D$ printed graphs, where maximum node numbers of around 30 nodes are common \cite{bae_computational_2024,drogemuller_haptic_2021}. 
Our flexible approach allows for the use of different sets of focus nodes as well, so that different aspects of the data can be examined with minimal reassembly, where, for $3D$ printed representations, printing times are often a limiting factor \cite{drogemuller_haptic_2021}.
Most of a \textit{HoloGraph} only has to be built once, and the various slices of a dynamic graph can be printed quickly and inexpensively. 
Compared to the approach of Bae et al.\cite{bae_computational_2024}, \textit{HoloGraphs} provides interactivity without the use of electronics or other augmentation.
The case study confirms that our \textit{HoloGraphs} can convey several facts about dynamic graph visualization at the hand of a simple example. 
A person with knowledge about the underlying data was able to quickly reason about important story events and character moments regarding several nodes of interest.
Thus, using a dataset already familiar to the user, made it possible to explain the mechanisms of dynamic networks to a person with no prior knowledge of or experience with dynamic graphs.
Finally, our domain expert also pointed out interesting aspects our representation could display: highlighting different groups in context slides could explain node movement in conjunction with group association, while the trajectory overlay could be easily adjusted to help find nodes at different time points. 

\paragraph{Limitations.} The number of timeslices that a \textit{HoloGraph} can display depends on several material factors. With the overhead projector slides we used, significant obstructions became apparent using about 20 slices (10 focus + 10 context slices). We used A5 pages (A4 halved) in landscape format for our embeddings. With a standard printer, sizes of up to an A4 page per timeslice are possible. Framing has to support larger slice formats well enough for the slices to remain parallel.
Drogemuller et al.~\cite{drogemuller_haptic_2021} show that users prefer haptic feedback in physical networks. Our approach, however, sacrifices the tactile component, in favor of visual clarity.

\paragraph{Future Work.} In this work, we investigated timesliced networks, however, 
many real-world networks are not discrete.
In future work, we aim to investigate the use of \textit{HoloGraphs} for continuous network physicalization.
Our evaluation presents some insights into how our method can help with understanding the underlying data, visualization techniques, and what insights can be extracted from dynamic network structures and behaviors. However, the versatility and accessibility of our method could support a multitude of dynamic data.



\bibliographystyle{apalike}
{\small
\bibliography{template}}



\end{document}